\documentstyle[]{article}
\title{ Low energy dynamics of $[U(1)]^{N}$ Chern-Simons solitons
               \thanks{Research supported by the grant KBN 2P302 049 05 } }
\author{Jacek Dziarmaga  \\
        Jagellonian University, Institute of Physics,  \\
        Reymonta 4,30-059 Krak\'ow, Poland
        \thanks{e-mail: ufjacekd@ztc386a.if.uj.edu.pl}}
\date{5 February 1994}
\begin{document}
\maketitle

    \begin{abstract}

    We apply the adiabatic approximation to investigate the low
energy dynamics of vortices in the parity invariant double self-dual Higgs
model with only mutual Chern-Simons interaction. When
distances between solitons are large they are particles subject to
the mutual interaction. The dual formulation of the model is derived
to explain the sign of the statistical interaction.
When vortices of different types pass one through another
they behave like charged particles in magnetic field. They can form
a bound state due to the mutual magnetic trapping. Vortices of the
same type exhibit no statistical interaction. Their short range interactions
are analysed. Possible quantum effects due to the finite width of vortices
are discussed.

    \end{abstract}

\section*{Introduction}

   Experiments with high temperature superconductors seem to show no indication
of parity breaking \cite{laughlin}. At the first sight this result seems
to exclude the anyonic mechanism of superconductivity. But it is not
the case as it was shown recently \cite{wilczek,hagen}. The presence of
Chern-Simons interaction
in a model does not lead inevitably to the breaking of the P and T invariance.
It just a property of the simplest models with only one Chern-Simons field.
When there is an even number of Chern-Simons fields and their coupling
constants are appropriately
choosen the parity invariance can be restored \cite{wilczek,hagen}.
One of the simplest models of this kind is the $[U(1)]^{2}$ model
of two Higgs fields each of them coupled to one of
the two Chern-Simons fields \cite{u1}. The model is constructed is such a way
that particles which carry one kind of charge interact with the magnetic
field of the other kind. So the ordinary construction of anyons as charged
particles which have at the same time attached a fictitious magnetic flux
is split into to parts. Particles of one kind are carriers of the charge
while those of the other kind carry the magnetic flux. Thus the ordinary
fractional statistics \cite{wilzee} is replaced by the so called mutual
statistics \cite{wilczek}.

   In this article we would like to investigate the Chern-Simons
interactions of vortices in the relativistic self-dual $[U(1)]^{N}$
model \cite{u1}. We apply Manton's idea of adiabatic approximation
\cite{manton}, with neccesary in the case of Chern-Simons vortices
corrections \cite{kimlee}, to
the topological solitons of the model. The topological solitons
configurations satisfy the lower Bogomolny bound on energy \cite{u1}
so the moduli space approximation is justified. Explicite calculations
are made in the special case of the $[U(1)]^{2}$ model but the results
can be easily generalised to the case of $[U(1)]^{N}$. At large separations
the vortices of different Higgs fields exhibit the expected mutual
interactions but with a sign opposite to that expected from their fluxes
and charges. It is very much like as for vortices in the standard $U(1)$
Chern-Simons-Higgs model which exhibit ordinary fractional statistics
properties \cite{kimmin,kimlee}. We derive dual formulation of the model
to explain the sign of the statistical interaction.

When the vortices pass one through another
their interaction
is a little more complicated. The pair of vortices of different types
behave like charged particles passing the flux of magnetic field
similarly as vortices in ordinary Chern-Simons-Higgs system \cite{kimlee,mis}.
Due to the fact that the spin of separated vortices
is equal to zero while that of coincident is nonzero there is a kind
of magnetic trapping - they form a composite state. If the corrections
to the standard adiabatic approximation are quantitative in nature
there is a periodic solution with vortices circling in the magnetic field
of the trap. On the other hand the vortices of the
same kind do not interact through the Chern-Simons field. If the
corrections to the ordinary adiabatic approximation amount only to the
renormalisation of parameters in effective Lagrangean, they would behave
very much like vortices in the Abelian Higgs model. In particular
the result of their head on collision would be the right angle scattering
\cite{ruback,samols}.

     The article is organised as follows. In Section 1 we derive the general
form of the effective Lagrangean. Section 2 is devoted to the long range
interactions of vortices and their mutual statistics. In the next
paragraph (3) we analyse what happens when various types of vortices pass
one through another. Section 4 is a presentation of the dual formulation of
the model. In the last section we summarise and discuss the results.

\section{ General form of the effective Lagrangean }

The Lagrangean of the relativistic model presented in \cite{u1} when we
restrict to such a choice of parameters that only mutual interactions
are preserved can be written in the form:
\begin{equation}\label{1.10}
  L=\kappa\varepsilon^{\mu\nu\lambda}v^{(1)}_{\mu}
                                  \partial_{\nu}v^{(2)}_{\lambda}+
    D_{\mu}\phi^{\star}D^{\mu}\phi+D_{\mu}\chi^{\star}D^{\mu}\chi-
    V(\phi,\chi) \;\; ,
\end{equation}
where the covariant derivatives are defined by
\begin{equation}\label{1.20}
  D_{\mu}\phi=\partial_{\mu}\phi-iq_{1}v^{(1)}_{\mu}\phi \;\; and \;\;
  D_{\mu}\chi=\partial_{\mu}\chi-iq_{2}v^{(2)}_{\mu}\chi \;\; ,
\end{equation}
while the Higgs potential of this self-dual model is equal to
\begin{equation}\label{1.25}
  V(\phi,\chi)=\frac{q^{2}_{1}q^{2}_{2}}{\kappa^{2}}
    [\mid\phi\mid^{2}(\mid\chi\mid^{2}-c^{2}_{2})^{2}+
     \mid\chi\mid^{2}(\mid\phi\mid^{2}-c^{2}_{1})^{2}]  \;\; .
\end{equation}
   First we will work out the general form of the effective Lagrangean
by a direct application of the methods of adiabatic approximation
known from the papers on slow motion of vortices in self-dual
Maxwell-Higgs system \cite{ruback,samols} and then we will discuss corrections
to this oversimplified version of the Lagrangean \cite{kimlee}.

   The Lagrangean (\ref{1.10}), when we eliminate auxillary $v_{0}^{(I)}$
components of the gauge fields with the help of the Gauss law
\begin{equation}\label{1.55}
  q_{1}v_{0}^{(1)}=\partial_{t}\omega_{1}
                   -\frac{\kappa B_{2}}{2q_{1}\mid\phi\mid^{2}} \;\;\;,\;\;\;
  q_{2}v_{0}^{(2)}=\partial_{t}\omega_{2}
                           -\frac{\kappa B_{1}}{2q_{2}\mid\chi\mid^{2}} \;\;,
\end{equation}
can be rewritten in the form
\begin{eqnarray}\label{1.60}
L&=&     -\kappa\varepsilon^{ij}v^{(1)}_{i}\partial_{t} v^{(2)}_{j}+
                 \frac{\kappa}{q_{1}}B_{(2)}\partial_{t}\omega_{1}+
                 \frac{\kappa}{q_{2}}B_{(1)}\partial_{t}\omega_{2}  \nonumber\\
 & &            +(\partial_{t}\mid\phi\mid)^{2}+
                 (\partial_{t}\mid\chi\mid)^{2}                     \nonumber\\
 & &          -\frac{\kappa^{2}B^{2}_{(2)}}{4q_{1}^{2}\mid\phi\mid^{2}}
                    -\frac{\kappa^{2}B^{2}_{(1)}}{4q_{2}^{2}\mid\chi\mid^{2}}
                    -\mid\vec{D}\phi\mid^{2}-\mid\vec{D}\chi\mid^{2}
                    -V(\phi,\chi) \;\; .
\end{eqnarray}
where we have introduced magnetic fields corresponding to the two
different gauge fields: $B_{I}=\varepsilon^{ij}\partial_{i}v_{j}$.
$\omega_{I}$-s are the phases of the Higgs fields. We have separated terms
containing time derivatives from those with spatial derivatives of the
fields. In this form of the Lagrangean we can make replacements due to the
identity
\begin{equation}\label{1.70}
  -\mid\vec{D}\phi\mid^{2}=
          -\mid(D_{1}+i\sigma_{1}D_{2})\phi\mid^{2}
          -\sigma_{1}q_{1}B_{(1)}\mid\phi\mid^{2}
          +i\sigma_{1}\varepsilon^{ij}\partial_{i}
                                         (\phi^{\star}D_{j}\phi -\;c.c.\;)
\end{equation}
and an analogous one for the field $\chi$.
The constants $\sigma_{1},\sigma_{2}$
which take values from the set $\{+1,-1\}$ have the same sign as the
topological indices
of field configurations $\phi$ and $\chi$ respectively.

    In the Manton's approximation we assume the fields to have the same form
as in some static solution and thus to fulfil the static field equations
at every instant of time. In our case the fields fulfil the self-dual
equations:
\begin{equation}\label{1.80}
  (D_{1}+i\sigma_{1}D_{2})\phi=0 \;\; and \;\;(D_{1}+i\sigma_{2}D_{2})\chi=0
\end{equation}
so the first term on the R.H.S. of eq.(\ref{1.70}) vanishes.
The assumed static solution depends in general on some finite set
of parameters which will play the role of collective coordinates.
To derive an effective Lagrangean we have to integrate out the spatial
dependence of the fields in eq.(\ref{1.60}). Assuming the Coulomb gauge
condition for the static solutions, which implies that the gauge potentials
can be written as
\begin{equation}\label{1.81}
   v^{(I)}_{i}=\varepsilon_{ij}\partial_{j}U^{(I)}
\end{equation}
with some regular functions $U^{(I)}$, we can remove first term in the
first line of (\ref{1.60}) and third term on RHS of Eq.(\ref{1.70}). Further
simplification can be achieved with a help of the remaining
static field equations
\begin{equation}\label{1.100}
  B_{(1)}=-2\sigma_{1}\frac{q_{1}q_{2}^{2}}{\kappa^{2}}
           \mid\chi\mid^{2}(\mid\phi\mid^{2}-c_{1}^{2}) \;\; ,
\end{equation}
\begin{equation}\label{1.110}
  B_{(2)}=-2\sigma_{2}\frac{q_{2}q_{1}^{2}}{\kappa^{2}}
           \mid\phi\mid^{2}(\mid\chi\mid^{2}-c_{2}^{2}) \;\; .
\end{equation}
Neglecting terms which are constant in a given topological sector
we obtain the following form of the effective Lagrangean
\begin{equation}\label{1.120}
   L^{(0)}_{eff}=\int\;d^{2}x\;
        [(\partial_{t}\mid\phi\mid)^{2}+(\partial_{t}\mid\chi\mid)^{2}+
        \frac{\kappa}{q_{1}}B_{2}\partial_{t}\omega_{1}+
        \frac{\kappa}{q_{2}}B_{1}\partial_{t}\omega_{2}]\;\;.
\end{equation}
   Y.Kim and K.Lee \cite{kimlee} tought us that it is important to take into
account corrections to such direct application of adiabatic approximation
which is oversimplified in the case of Chern-Simons vortices. Namely
even in the slow motion approximation the fields can not be taken just
as static solutions with time dependent parameters. We also have to take
into account corrections to those fields which are linear in velocities.
   \begin{equation}\label{1.135}
     \phi\rightarrow \phi+\delta\phi \;\;\;,\;\;\;
     \chi\rightarrow \chi+\delta\chi \;\;\;,\;\;\;
     v^{(I)}_{\mu}\rightarrow v^{(I)}_{\mu}+\delta v^{(I)}_{\mu}
   \end{equation}
because they give additional terms to the effective Lagrangean. It is a kind
of complicated Lorenz transformation.

  We can show that corrections affect only this part of effective Lagrangean
which is quadratic in velocities. Let us take closer look at the Lagrangean
in the form of Eq.(\ref{1.60}). Second
line of the above formula is manifestly of at least second order in
velocities. Third line is minus static energy density, so the static
solutions are its stationary points. Thus third line also gives
only quadratic terms. The only contribution to the linear part
can come from the first line. Any replacements of the form (\ref{1.135})
will produce only extra second or higher order terms. So we have to take
the first line as it stands - only with static fields with time-dependent
parameters. As it has been  shown the first term of the first line gives no
contribution, so the linear part of the Lagrangean remains as
\begin{equation}\label{app.30}
  L^{(1)}_{eff}=\kappa\int d^{2}x[\frac{B_{(2)}}{q_{1}}\partial_{t}\omega_{1}
                    +\frac{B_{(1)}}{q_{2}}\partial_{t}\omega_{2}] \;\;.
\end{equation}
Another representation of $L^{(1)}_{eff}$, useful for further discussion
can be obtained with use of the identity
\begin{equation}\label{13.1}
  \dot{\omega}_{I}=
    \sigma_{I}\frac{d}{dt}\sum_{p_{I}}Arg(\vec{x}-\vec{R}_{p_{I}})=
    \sigma_{I}\sum_{p_{I}}\varepsilon_{ij}\dot{R}^{i}_{p_{I}}
                      \partial_{j}\ln\mid\vec{x}-\vec{R}_{p_{I}}\mid  \;\;
\end{equation}
and certain integration by parts
\begin{equation}\label{13.2}
  L^{(1)}_{eff}=\frac{2\pi\kappa\sigma_{1}}{q_{1}}\sum_{p_{1}}
                \dot{R}^{i}_{p_{1}}v^{(2)}_{i}(\vec{R}_{p_{1}})+
                \;\;\; (\;1\leftrightarrow 2\;)  \;\;.
\end{equation}
$\vec{R}_{p_{I}}(t)$ are positions of vortices of type "I",I=1,2. From this
representation one can obtain general form of the orbital part of angular
momentum
\begin{eqnarray}\label{13.3}
  J_{orb}&=&\sum_{I}\sum_{p_{I}}\varepsilon_{ij}R^{i}_{p_{I}}
         \frac{\partial L^{1}_{eff}}{\partial\dot{R}^{j}_{p_I}}=  \nonumber\\
  & &2\pi\kappa[\frac{\sigma_{1}}{q_{1}}\sum_{p_{1}}R^{i}_{p_{1}}
                \varepsilon_{ij}v^{(2)}_{j}(\vec{R}_{p_{1}})+
                \;\;(\;1\leftrightarrow 2\;)\;]                \;\;.
\end{eqnarray}
The form of $L^{(2)}_{eff}$ can be extracted from (\ref{1.60}) after
replacement (\ref{1.135}) and making use of static equations fulfiled by
"static" fields. It gives correction to the effective Lagrangean (\ref{1.120})
\begin{equation}\label{app.40}
  \delta L_{eff}[fields,\delta.fields]
\end{equation}
which is a functional of "static" fields, their linear corrections and
their time derivatives. It is a long expression which we will not write down.
To make use of this expression we have to express "$\delta$-fields" in terms
of "fields" and/or positions of vortices. We have to know corrections
for a given trajectory in the parameter space. We can make replacement
(\ref{1.135}) in the field equations following from (\ref{1.10})
\begin{equation}\label{app.50}
  D_{\mu}D^{\mu}\phi=-\frac{\partial V}{\partial\phi^{\star}}\;\;\;,\;\;\;
  \kappa\varepsilon^{\mu\nu\lambda}\partial_{\nu}v_{\lambda}^{(2)}=
  iq_{1}(\phi D^{\mu}\phi^{\star}-c.c.)
\end{equation}
and analogous one for the field $\chi$. It is convenient to use geodesic
parametrisation on moduli space. By geodesic parametrisation we mean such
a set of parameters that during the time evolution of the field configuration
their accelerations are vanishing. It is always in principle possible
to construct such geodesic coordinates at least for a finite period of time.
One advantage of such a set of coordinates is that we can neglect in
Eqs.(\ref{app.50}) terms which contain accelerations or higher order time
derivatives of parameters. The other one also very important is that
only when velocities in a given parametrisation are constant
there is a direct correspondence between the fact that kinetic energy is small
and smallness of velocities. For example in the head-on collision
of Nielsen-Olesen vortices the time derivatives of cartesian coordinates
of vortices become singular during collision \cite{ruback,samols},
so in this case terms quadratic in such velocities are much larger then linear
terms and obviously can not be neglected. Nethertheles the kinetic energy
is still small - it is still a slow motion. In fact it is possible to
find geodesic coordinates good for small separations of these vortices.
So only for geodesic parametrisation it is reliable to preserve
in Eqs.(\ref{app.50}) only terms which are linear in velocities and neglect
accelerations and higher order time derivatives.

   Thus "simplified" equations read
\begin{eqnarray}\label{app.60}
0&=&\partial_{k}^{2}\delta f_{1}+
[q_{1}^{2}v_{0}^{(1)}v_{0}^{(1)}
-(\partial_{k}\omega_{1}-q_{1}v^{(1)}_{k})^{2}]\delta f_{1}+
2q_{1}f_{1}(\partial_{k}\omega_{1}-q_{1}v^{(1)}_{k})\delta v^{(1)}_{k}
                                                                   \nonumber\\
 & &-2q_{1}f_{1}v^{(1)}_{0}(\dot{\omega}_{(1)}-q_{1}\delta v_{0}^{(1)})-
\frac{\partial^{2}V}{\partial f_{1}^{2}}\delta f_{1}-
\frac{\partial^{2}V}{\partial f_{1}\partial f_{2}}\delta f_{2}     \nonumber\\
0&=&  f_{1}(\dot{v}_{0}^{(1)}+\partial_{k}\delta v^{k}_{(1)})+
                                    2q_{1}v_{0}^{(1)}\dot{f}_{1}   \nonumber\\
0&=&  \kappa\varepsilon_{ij}\partial_{i}\delta v_{j}^{(2)}-
q_{1}f_{1}^{2}(\dot{\omega}_{1}-q_{1}\delta v_{0}^{(1)})+
2q_{1}^{2}v_{0}^{(1)}f_{1}\delta f_{1}                             \nonumber\\
0&=&  \kappa\varepsilon_{ij}(\partial_{j}\delta v_{0}^{(2)}-\dot{v}^{(2)}_{j})+
q_{1}^{2}f_{1}^{2}\delta v^{(1)}_{k}-
2q_{1}f_{1}(\partial_{k}\omega_{1}-q_{1}v_{k}^{(1)})\delta f_{1}   \;\;
\end{eqnarray}
and an analogous set of equations for the field $\chi$. We have introduced
moduli and phases of the Higgs fields:
$\phi=\frac{f_{1}}{\sqrt{2}}e^{i\omega_{1}}$ ,
$\chi=\frac{f_{2}}{\sqrt{2}}e^{i\omega_{2}}$.
An analogous set of equations was derived in \cite{kimlee} for self-dual
Chern-Simons-Higgs model. Neglecting of second order
time derivatives for any kind of coordinates can lead to serious problems.
For example in the papers on effective string models for Nielsen-Olesen
vortices the second order derivatives on world-sheet parameters
were neglected. It was shown in \cite{arodz1} that it was the reason why the
string with rigidity was obtained, which is known to posses classical
tachionic solutions \cite{arodz2}.

   With a help of Eqs.(\ref{app.60}) $\delta L_{eff}$ can be simplified
to the following form
\begin{eqnarray}\label{app.70}
 \delta L_{eff}=\sum_{I}\int d^{2}x\;
    \frac{1}{2}f_{I}^{2}[(\dot{\omega}_{I}-q_{I}v_{0}^{I})^{2}+
                                          q_{I}^{2}(\delta v_{i})^{2}]
 -f_{I}\delta f_{I}q_{I}v_{o}^{(I)}(\dot{\omega}_{I}-q_{I}v_{0}^{(I)})
                                                                  &\nonumber \\
 -f_{I}\delta f_{I}(\partial_{i}\omega_{I}-q_{I}v_{i}^{(I)})
                                           q_{I}\delta v_{i}^{(I)}& \;\;.
\end{eqnarray}
Once we have solved Eqs.(\ref{app.60}) we can substitute their solutions
to the above functional and integrate out their spatial dependence. We will
be left with a mechanical Lagrangean which should be reparametrisation
invariant. From that point on we will be able to use any parametrisation
we like. But to evaluate $\delta L_{eff}$ we have first to guess
geodesic coordinates and then to solve Eqs.(\ref{app.60}). We have to try
with different parametrisations and then to check whether for a given
parametrisation there exist solution to Eqs(\ref{app.60}). If yes than
as a matter of fact we have found already the trajectory. The effective
Lagrangean can be evaluated for consistency check and because it is useful
if we want to perform effective quantisation of the theory or to investigate
its effective thermodynamics \cite{shah}. The effective Lagrangean under
restrictions due to the above comments reads
  \begin{equation}\label{app.80}
    L_{eff}=\int d^{2}x\;[(\partial_{t}\mid\phi\mid)^{2}+
            (\partial_{t}\mid\chi\mid)^{2}]+\delta L_{eff}+
            \kappa\int d^{2}x\;[\frac{B_{2}}{q_{1}}\dot{\omega}_{1}+
                                \frac{B_{1}}{q_{2}}\dot{\omega}_{2}] \;\;.
  \end{equation}
  To start the above procedure we have to make some "educated" guesswork.
To provide an appropriate basis for it in the next two sections we investigate
ordinary adiabatic approximation in detail. Because of mathematical
difficulties (existence proof and/or explicite solution of Eqs.(\ref{app.60}))
we postpone calculation of corrections to future publication.

\section{ Long range interactions of solitons }

  For sufficiently separated vortices we can approximate gauge invariant
fields: $(\partial_{k}\omega_{I}-q_{I}v_{k}^{I})$ by contributions
due to particular vortices. At the core of any choosen vortex such fields
due to the other vortices are very small - they vanish exponentially with
distances. Since
$\omega_{I}=\sigma_{I}\sum_{p_{I}}Arg(\vec{x}-\vec{R}_{p_{I}})$
is an exact solution of static equations for any configuration of vortices,
corrections to the simple superposition of gauge potentials are negligible.
Thus values of the gauge fields in Eq.(\ref{13.2}) can be obtained from
a formula
\begin{equation}
  v^{(2)}_{i}(\vec{R}_{p_{1}})=-\frac{\sigma_{2}}{q_{2}}\sum_{q_{1}}
                    \frac{\varepsilon_{ij}(R_{p_{1}}-R_{q_{2}})^{j}}
                        {\mid\vec{R}_{p_{1}}-\vec{R}_{q_{2}}\mid^{2}} \;\;,\;\;
                                            (\;1\leftrightarrow 2\;)
\end{equation}
so the linear part of effective Lagrangean is
\begin{equation}\label{app.81}
  L^{(1)}_{eff}=2\pi\kappa\frac{\sigma_{1}\sigma_{2}}{q_{1}q_{2}}
                    \sum^{n_{1}}_{p_{1}=1} \sum^{n_{2}}_{p_{2}=1}
                    \frac{d}{dt}
                    Arg( \vec{R}^{(1)}_{p_{1}} - \vec{R}^{(2)}_{p_{2}} ) \;\;.
\end{equation}
The Lagrangean contains only terms of mutual statistical interaction
between vortices of different types.

To obtain the kinetic term one has to evaluate the first term of the
effective lagrangean (\ref{app.80}) and $\delta L_{eff}$. We approximate the
moduli of the Higgs
fields by a normalised product of the fields of isolated unit vortices:
\begin{equation}\label{2.70}
  \mid\phi(\vec{x})\mid=c_{1}\prod^{n_{1}}_{p_{1}=1}
                         G(\mid \vec{x} - \vec{R}^{(1)}_{p_{1}} \mid) \;\;,
\end{equation}
\begin{equation}\label{2.80}
  \mid\chi(\vec{x})\mid=c_{2}\prod^{n_{2}}_{p_{2}=1}
                         G(\mid \vec{x} - \vec{R}^{(2)}_{p_{2}} \mid) \;\;,
\end{equation}
where $G$ is a profile of a unit vortex the same for the two types,
which satisfies the equation
\begin{equation}\label{2.90}
  \nabla^{2}\ln G^{2}=
     \frac{4q_{1}^{2}q_{2}^{2}c_{1}^{2}c_{2}^{2}}{\kappa^{2}}(G^{2}-1) \;\;,
\end{equation}
with the boundary conditions: $G(0)=0$ and $G(\infty)=1$.
The quadratic part of the effective Lagrangean reads
\begin{equation}\label{2.100}
  L^{(2)}_{eff}=\frac{1}{2} \bar{M}
                   [ c_{1}^{2} \sum_{p_{1}=1}^{n_{1}} \vec{V}^{2}_{p_{1}} +
                     c_{2}^{2} \sum_{p_{2}=1}^{n_{2}} \vec{V}^{2}_{p_{2}} ]
                +\delta L_{eff} \;\;,
\end{equation}
where the coefficient $\bar{M}$ equals to
\begin{equation}\label{2.110}
  \bar{M}=2\pi\int rdr\;(\frac{dG(r)}{dr})^{2}     \;\;.
\end{equation}
About $\delta L_{eff}$ we know only that it is quadratic in velocities.
Let us restrict to the case of two vortices of the same type and choose
the center of mass frame
\begin{equation}\label{2.112}
  L^{(2)}_{eff}=g_{ij}(R^{k})\dot{R}^{i}\dot{R}^{j}\;\;,\;\;g_{ij}=g_{ji}\;\;,
\end{equation}
where $R_{k}$-s are coordinates of the choosen vortex. Rotational invariance
restricts this form to
\begin{equation}\label{2.114}
  L^{(2)}_{eff}=g_{1}(R)\dot{R}^{2}+g_{2}(R)R^{2}\dot{\Theta}^{2} \;\;,
\end{equation}
where $R^{1}+iR^{2}=Re^{i\Theta}$. At very large $R$ we expect the
influence of one vortex on another to be very small:
$g_{1}(R)\;,\;g_{2}(R)\rightarrow const$ as $R\rightarrow\infty$. This
reasoning can be repeated for any pair of vortices. Finally we obtain
\begin{equation}\label{2.120}
  L_{eff}=\frac{1}{2} M
                   [ c_{1}^{2} \sum_{p_{1}=1}^{n_{1}} \vec{V}^{2}_{p_{1}} +
                     c_{2}^{2} \sum_{p_{2}=1}^{n_{2}} \vec{V}^{2}_{p_{2}} ]+
                 2\pi\kappa\frac{\sigma_{1}\sigma_{2}}{q_{1}q_{2}}
       \sum^{n_{1}}_{p_{1}=1} \sum^{n_{2}}_{p_{2}=1}
       \frac{d}{dt}Arg( \vec{R}^{(1)}_{p_{1}} - \vec{R}^{(2)}_{p_{2}} ) \;\;.
\end{equation}
where $M=\bar{M}+\delta M$ is the effective mass with included corrections
from $\delta L_{eff}$. Corrections can be calculated with a help of the
formula (\ref{app.70}), since for fairly separated vortices
"$\delta$-fields" are given by ordinary Lorenz formulas linearised in
velocities. The corrected coefficient $M$ appears to be equal $2\pi$, what
is consistent with the original field theoretical model since
the energy of a static unit vortex of type "I" is equal to $2\pi c_{I}^{2}$.

   We can see that when the widths of the vortices can be neglected as
compared with their separations the system behaves like a set of free
particles with mutual statistical interactions, at least in the slow
motion approximation. What happens if the vortices come into
very close encounters of one another is a subject of the next section.

Let us remark here on the possibility of ordinary fractional statistics in
the system if the short range interactions between vortices of different
types fauvored them to form mixed anyonic $\phi-\chi$ bound states.

\section{ Short range interactions }

In this section we would like to investigate interactions of the two types
of vortices when their cores overlap. To apply the Manton's prescription
we have to know at least an approximate static solution. Let us take as
a zero order approximation the configuration of vortices sitting on top
of each other and
then let us find a small perturbation of the field dependent on a definite set
of parameters.

 Let the topological
indices of the fields $\phi$ and $\chi$ be $\sigma_{1}n_{1}$ and
$\sigma_{2}n_{2}$ recpectively, where $n$-s are positive integers and
$\sigma$-s take values $+1$ or $-1$. The solution corresponding to
vortices sitting on top of each other takes the form:
\begin{equation}\label{3.10}
  \phi=c_{1}F(r)e^{i\sigma_{1}n_{1}\theta}\;\;,\;\;
  \chi=c_{2}H(r)e^{i\sigma_{2}n_{2}\theta}\;\;,
\end{equation}
\begin{equation}\label{3.20}
  \vec{v}_{(1)}=\vec{e}_{\theta}\sigma_{1}V^{\theta}_{(1)}\;\;,\;\;
  \vec{v}_{(2)}=\vec{e}_{\theta}\sigma_{2}V^{\theta}_{(2)}\;\;.
\end{equation}
Upon substitution of the above Ansatz to the static field equations
(\ref{1.80},\ref{1.100},\ref{1.110}) one obtains
\begin{equation}\label{3.30}
  F'-\frac{n_{1}}{r}F+q_{1}FV^{(1)}_{\theta}=0 \;\;,
\end{equation}
\begin{equation}\label{3.40}
  H'-\frac{n_{2}}{r}H+q_{2}HV^{(2)}_{\theta}=0 \;\;,
\end{equation}
\begin{equation}\label{3.50}
  \frac{-1}{r}\frac{\partial(rV^{(1)}_{\theta})}{\partial r}=
                                        q_{2}\gamma H^{2}(F^{2}-1) \;\;,
\end{equation}
\begin{equation}\label{3.60}
  \frac{-1}{r}\frac{\partial(rV^{(2)}_{\theta})}{\partial r}=
                                        q_{1}\gamma F^{2}(H^{2}-1) \;\;,
              \;\;\gamma=\frac{2q_{1}q_{2}c_{1}^{2}c_{2}^{2}}{\kappa^{2}}\;\;.
\end{equation}
This zero order solution together with a small perturbation reads
\begin{equation}\label{3.70}
  \phi=c_{1}F(r)[1+f(r,\theta)]
                    e^{i\sigma_{1}n_{1}\theta+i\alpha_{1}(r,\theta)} \;\;,
\end{equation}
\begin{equation}\label{3.80}
  \chi=c_{2}H(r)[1+h(r,\theta)]
                    e^{i\sigma_{2}n_{2}\theta+i\alpha_{2}(r,\theta)} \;\;,
\end{equation}
\begin{equation}\label{3.90}
  \vec{v}_{(I)}=\vec{e}_{\theta}\sigma_{I}V_{(I)}^{\theta}+\vec{a}_{(I)}\;\;,
                                                       \;\; I=1,2 \;\;
\end{equation}
Upon substitution of the above form of the solution to
the self-dual equations (\ref{1.80}) and subsequent linearisation one obtains
the following first order equations
\begin{equation}\label{3.100}
  q_{I}a^{(I)}_{\theta}=\frac{1}{r}\partial_{\theta}\alpha^{(I)}
                                     -\sigma_{I}\partial_{r}f^{(I)} \;\;,
\end{equation}
\begin{equation}\label{3.110}
  q_{I}a^{(I)}_{r}=\partial_{r}\alpha^{(I)}
                      +\sigma_{I}\frac{\partial_{\theta}f^{(I)}}{r} \;\;.
\end{equation}
These equations enable us to find perturbations of the gauge fields
once $f^{(I)}=(f,h)$ and $\alpha_{(I)}$-s are already known.

  After substitution of eqs.(\ref{3.90}) to the Coulomb gauge
condition, $\partial_{i}v^{(I)}_{i}=0$, linearisation and use of
eqs.(\ref{3.100},\ref{3.110}) an equation determining the phases of the
Higgs fields appears:
\begin{equation}\label{3.120}
  \nabla^{2}\alpha_{(I)}=0 \;\;.
\end{equation}
 Similarly equations (\ref{3.30},\ref{3.40}) linearised
in the perturbations yield
\begin{equation}\label{3.130}
  \nabla^{2}f=\bar{\gamma}H^{2}[F^{2}f+(F^{2}-1)h] \;\; ,
\end{equation}
\begin{equation}\label{3.140}
  \nabla^{2}h=\bar{\gamma}F^{2}[H^{2}h+(H^{2}-1)f] \;\; ,
\end{equation}
where $\bar{\gamma}=2q_{1}q_{2}\gamma$.
{}From the two above equations one can obtain general form of the pertubation
of the moduli of the Higgs fields. Then one have to choose such
a solution of eq.(\ref{3.120}) as to avoid singularities of the gauge fields,
see eqs.(\ref{3.100},\ref{3.110}). We will solve this problem explicitely
in two special cases.

\subsection{ Interaction of unit $\phi$-vortex with unit $\chi$-vortex. }

In this case we have unit topological indices of both the $\phi$-field
configuration and the $\chi$-field one
\begin{equation}\label{i.10}
   n_{1}=1\;\;,\;\;n_{2}=1\;\;.
\end{equation}
  Eqs.(\ref{3.30}-\ref{3.60}) can be simplified by a substitution
\begin{equation}\label{i.20}
  F(r)=H(r)\equiv G(r) \;\;,\;\;
  \frac{V^{(1)}_{\theta}}{q_{2}}=
  \frac{V^{(2)}_{\theta}}{q_{1}}\equiv V_{\theta}(r) \;\;,
\end{equation}
  to the form
\begin{equation}\label{i.30}
  G'-\frac{G}{r}+q_{1}q_{2}GV_{\theta}=0 \;\;,
\end{equation}
\begin{equation}\label{i.40}
  \frac{-1}{r}\frac{\partial(rV_{\theta})}{\partial r}=
                                          \gamma G^{2}(G^{2}-1) \;\;.
\end{equation}
  From the index theorem \cite{u1} we know that in the special case of
$n_{1}=n_{2}=1$ there are only two splitting modes for each of the two
types of the fields. We can Fourier-transform $f$ and $h$ in $\theta$:
\begin{equation}\label{i.50}
  f(r,\theta)=
       f(r)[\lambda_{1}cos(\sigma_{1}\theta)
                       +\lambda_{2}sin(\sigma_{1}\theta)] \;\;,
\end{equation}
\begin{equation}\label{i.60}
  h(r,\theta)=
       h(r)[\mu_{1}cos(\sigma_{2}\theta)+\mu_{2}sin(\sigma_{2}\theta)] \;\;
\end{equation}
The $\theta$-independent
terms are neglected because solitons have a definite size, while the
terms higher than the first would spoil regularity of the total
Higgs fields. To define the meaning of coefficiens $\lambda,\mu$ we normalise
the radial functions in such a way that
\begin{equation}\label{i.70}
  f(r)\sim\frac{-1}{r}\;,\;h(r)\sim\frac{-1}{r}\;,\;as\;r\rightarrow0 \;\;.
\end{equation}
Eventual higher order terms would have a stronger singularity which
could not be matched by $G(r)\sim r$, see eqs.(\ref{3.70},\ref{3.80}).
Thus eqs.(\ref{i.60},\ref{i.70}) present the general form of the perturbation
compatible with the regularity of Higgs fields.

  To avoid singularities in the gauge fields, eqs.(\ref{3.100},\ref{3.110}),
we take the perturbations of phases
\begin{equation}\label{i.120}
  \alpha_{1}=\frac{-1}{r}[\lambda_{2}cos(\sigma_{1}\theta)
                                    -\lambda_{1}sin(\sigma_{2}\theta)] \;\;,
\end{equation}
\begin{equation}\label{i.130}
  \alpha_{2}=\frac{-1}{r}[\mu_{2}cos(\sigma_{2}\theta)
                                   -\mu_{1}sin(\sigma_{2}\theta)] \;\;,
\end{equation}
which satisfy eqs.(\ref{3.120}). Now we can see that the perturbed
Higgs fields in the limit of small $r$ are proportional to
\begin{equation}\label{i.140}
  \phi\simeq(z_{\sigma_{1}}-\lambda) \;\;,
          \;\;\chi\simeq(z_{\sigma_{2}}-\mu)\;\;,
\end{equation}
where we have introduced $z_{\sigma_{I}}=x+i\sigma_{I}y$ and
$\lambda=\lambda_{1}+i\lambda_{2}\;,\;\mu=\mu_{1}+i\mu_{2}$.
The effect of the perturbation is a shift of the zeros of the Higgs
fields to $\lambda_{\sigma_{1}}$ and to $\mu_{\sigma_{2}}$,
up to linear terms. To work in the center of mass frame we have to choose
$\mu_{\sigma_{2}}=-\lambda_{\sigma_{1}}$ together with $c_{1}=c_{2}=c$.
Last condition can be suspended if we make certain relative rescaling
of the parameters $\lambda$ and $\mu$. Without loose of generality in
evaluation of $f(r)$ and $h(r)$ we can choose $\lambda_{2}=\mu_{2}=0$.
Now upon substitution of eqs.(\ref{i.50},\ref{i.60}) to
eqs.(\ref{3.130},\ref{3.140}) one obtains
\begin{equation}\label{i.80}
  \triangle_{1}f(r)=\bar{\gamma}G^{2}[G^{2}f(r)-(G^{2}-1)h(r)] \;\;,
\end{equation}
\begin{equation}\label{i.90}
  \triangle_{1}h(r)=\bar{\gamma}G^{2}[G^{2}h(r)-(G^{2}-1)f(r)] \;\;,
  \;\;\triangle_{k}\equiv
      (\frac{d^{2}}{dr^{2}}+\frac{1}{r}\frac{d}{dr}-\frac{k^{2}}{r^{2}}) \;\;.
\end{equation}
We can make the replacements $f(r)=w(r)+u(r)$ and $h(r)=w(r)-u(r)$,
where the newly introduced functions fulfil the equations
\begin{equation}\label{i.100}
    \triangle_{1}w(r)=\bar{\gamma}G^{2}w(r) \;\;,
\end{equation}
\begin{equation}\label{i.110}
    \triangle_{1}u(r)=\bar{\gamma}G^{2}(2G^{2}-1)u(r) \;\;.
\end{equation}
The only solution compatible with normalisation (\ref{i.70}) is $u(r)=0$ and
$f(r)=h(r)=w(r)$. $w(r)$ can be approximated for large $r$ by the
modified Bessel function
$K_{1}(\rho) \sim \sqrt{\frac{\pi}{2\rho}}e^{-\rho}
                                 \;\;,\rho\equiv\sqrt{\bar{\gamma}}r$.
Going from infinity to zero this approximate solution is replaced
by a linear combination $\delta_{+1}r+\delta_{-1}\frac{1}{r}$. In the case
of $\delta_{-1}\neq 0$, which we think to be quite general,
it is possible to rescale the whole function $w(r)$ in such a way that
we obtain an assymptotics of eq.(\ref{i.70}). The perturbations
of the Higgs fields are square-integrable.

  Now the effective Lagrangean (\ref{app.80}) reads
\begin{equation}\label{i.150}
  L_{eff}=\frac{1}{2}M_{eff}(\dot{R}^{2}+R^{2}\dot{\Theta}^{2})
                      +\delta L_{eff}
                      +B_{eff}R^{2}\dot{\Theta} \;\;.
\end{equation}
$R$ and $\Theta$ are the polar coordinates of the zero of the $\phi$-field:
$\lambda\equiv Re^{i\Theta}$, while the introduced coefficients are
an effective reduced mass of the two vortices
\begin{equation}\label{i.160}
  M_{eff}=4\pi c^{2} \int_{0}^{\infty} (rdr) G^{2}(r) w^{2}(r) \;\;,
\end{equation}
and an effective "uniform external magnetic field"
\begin{equation}\label{i.170}
  B_{eff}=\frac{8\pi q_{1}q_{2}c^{4}}{\kappa\sigma_{1}\sigma_{2}}
                                    \int_{0}^{\infty} dr G^{2}(r)[-w(r)] \;\;.
\end{equation}
Let us first discuss linear part of the Lagrangean (\ref{i.150}), which becomes
exact for very small $R$. It is a term which describes, as it stands, coupling
of a charged particle to a uniform external magnetic field perpendicular to
the plane. The total angular momentum in the effective description
is (for $\dot{\Theta}=0$, $R\rightarrow 0$):
\begin{equation}\label{i.180}
  J(R)\stackrel{def}{=}-\frac{2\pi\kappa\sigma_{1}\sigma_{2}}{q_{1}q_{2}}+
                     \frac{\partial L_{eff}}{\partial \dot{\Theta}}=
                       -\frac{2\pi\kappa\sigma_{1}\sigma_{2}}{q_{1}q_{2}}+
                                                B_{eff}R^{2}+\;O(R^{3}) \;\;,
\end{equation}
where we have shifted the scale so that for $R\rightarrow 0$, J tends to the
value of spin characteristic for coincident static unit $\phi$ and $\chi$
vortices. With this choice of the scale we obtain from (\ref{2.120})
that for large $R$: $J\rightarrow 0$. This result is consistent with what we
know from field-theoretical considerations. Separate $\phi$ or $\chi$ mutually
interacting vortices carry no spin. Spin is nonzero only when their cores
overlap. Eq. (\ref{i.180}) gives leading terms in expansion of $J(R)$ around
$R=0$.

  It is interesting that Eq.(\ref{i.180}) can be inverted in a remarkable way
\begin{equation}\label{ins.10}
  B_{eff}=\lim_{R\rightarrow 0}\frac{1}{R}(\frac{J(R)-J(0)}{R})=
          \lim_{R\rightarrow 0}\frac{1}{R}\frac{dJ(R)}{dR}  \;\;.
\end{equation}
Natural thing is to ask whether such a formula can be generalised to arbitrary
value of $R$. Let us look at $L^{(1)}_{eff}$ in the form of Eq.(\ref{13.2})
\begin{equation}\label{ins.20}
  L^{(1)}_{eff}=\frac{2\pi\kappa\sigma_{1}}{q_{1}}\sum_{p_{1}}
                          \dot{R}^{i}_{p_{1}}v_{i}^{(2)}(\vec{R}_{p_{1}})+
                                         \;\;(\;1\leftrightarrow 2\;) \;\;.
\end{equation}
 What we see is an interaction term which couples point particle currents
to fields $v_{i}^{(I)}$ defined on the moduli space. Due to this coupling
vortex at $\vec{R}_{p_{1}}$ feels magnetic field
\begin{equation}\label{ins.30}
  B^{(1)}_{eff}(\vec{R}_{p_{1}})=\frac{2\pi\kappa\sigma_{1}}{q_{1}}
             \varepsilon_{ij}\partial_{i}v_{j}^{(2)}(\vec{R}_{p_{1}})\;\;,\;\;
             \partial_{i}=\frac{\partial}{\partial R^{i}_{p_{1}}} \;\;.
\end{equation}
Our pair of vortices feels double this field:
$B_{eff}=B^{(1)}_{eff}+B^{(2)}_{eff}$. From the formula (\ref{13.3})
we obtain angular momentum, which in our case reads
\begin{equation}\label{ins.40}
  J_{orb}(\vec{R})=2\pi\kappa\frac{\sigma_{1}}{q_{1}}R^{i}
                                     \varepsilon_{ij}v_{j}^{(2)}(\vec{R})
                  -2\pi\kappa\frac{\sigma_{2}}{q_{2}}R^{i}
                                    \varepsilon_{ij}v_{j}^{(1)}(-\vec{R})\;\;,
\end{equation}
 Due to the rotational symmetry, in polar coordinates
\begin{equation}\label{ins.50}
  J_{orb}(R)=2\pi\kappa\frac{\sigma_{1}}{q_{1}}Rv_{\theta}^{(2)}(R)+
             2\pi\kappa\frac{\sigma_{2}}{q_{2}}Rv_{\theta}^{(1)}(R) \;\;.
\end{equation}
Now we can see that
\begin{eqnarray}\label{ins.60}
  \frac{1}{R}\frac{dJ}{dR}=\frac{1}{R}\frac{dJ_{orb}}{dR}=
   2\pi\kappa\frac{\sigma_{1}}{q_{1}}
   (\frac{v_{\theta}^{(2)}}{R}+\frac{dv_{\theta}^{(2)}}{dR})+
   \;\;(\;1\leftrightarrow 2\;)=&                              \nonumber\\
   =B^{(1)}_{eff}(R)+B^{(2)}_{eff}(R)=B_{eff}(R)& \;\;
\end{eqnarray}
Thus eq.(\ref{ins.10}) can indeed be generalised for any value of $R$:
\begin{equation}\label{ins.70}
   B_{eff}(R)=\frac{1}{R}\frac{dJ(R)}{dR} \;\;,
\end{equation}
where $B_{eff}(R)$ is a magnetic field felt by a reduced particle in our
problem, while $J(R)$ is a total field-theoretical spin, which can be obtained
numerically. It is important that this formula is based on global properties
of the field configurations and not on local distortions of the fields,
so $B_{eff}(R)$ can be calculated numerically with great accuracy.

  Just from the knowledge of spin dependence on $R$ we can obtain
qualitative understanding of interactions between $\phi$ and $\chi$ vortices.
Vortices boosted against each other for a head-on collision will avoid
direct collision. Their initial total angular momentum is zero. For very
small separations spin itself would have to be close to
$(-\frac{2\pi\kappa\sigma_{1}\sigma_{2}}{q_{1}q_{2}})$ so the vortices
must aquire angular momentum of the opposite value. For
$\frac{\sigma_{1}\sigma_{2}}{q_{1}q_{2}}$ positive they will be turned
by $B_{eff}(R)$ to the right while for
the negative value of the coefficient they
will turn left. Since $B_{eff}(R)$ is short ranged, for sufficiently small
initial velocity it becomes impossible to reach the center of mass for any
value of impact parameter. Analogous argument shows that if a $\phi$ vortex
initially sits on top of $\chi$ vortex there is a velocity small enough
below which they can not escape to infinity. So it is a kind of magnetic
trap for vortices. They trap each other and form a composite. From
Eq.(\ref{2.120})
is clear that such composites as a whole behave like anyons. Their internal
reduced dynamics is that of a particle in an external magnetic field. Upon
quantisation it is probable to obtain resonances classified
by the Landau levels. The trapped vortices do
not form a real bound state - their energy is the same as that of isolated
vortices.

     Now let us take a look at the quadratic part of the effective Lagrangean.
If we assume that $\delta L_{eff}$ only renormalises $M_{eff}$ than $L_{eff}$
is a Lagrangean for a planar motion of charged particle in external uniform
magnetic field perpendicular to the plane. Solutions to the equations of
motion are circular trajectories
\begin{equation}\label{i.200}
  R(t)e^{i\Theta(t)}=A+Be^{-i\Omega t} \;\;,\;\;
  \Omega=\frac{2 B_{eff}}{M_{eff}} \;\;,
\end{equation}
where $A,B$ are complex constants. This solution is valid only for small $R$.
The trapped vortices rotate around circles (not necessarily around the center
of mass if $A\neq 0$). We can also obtain qualitative understanding of
scattering. Let us assume that this approximation is valid up to say
$R_{0}$, for larger $R$ let the vortices move along straight lines. In the
head-on collision vortices cross the circle $R=R_{0}$, move along an arc of
a circle and escape to infinity. For larger velocities, if the adiabatic
approximation still works, scattering pattern evolves to a forward
scattering. We can check whether $M_{eff}$ is indeed a constant for small
$R$ by performing numerical simulation of the above described head-on
collision. For example if for higher velocities right-angle scattering
is obtained, $M_{eff}$ must behave like $R^{2}$ for small $R$. From
analogous simulations for Nielsen-Olesen vortices we can expect
that the adiabatic approximation can still work well even up to
$\frac{1}{3}$ of the light velocity.

  The eventual dependence of $M_{eff}$ on $R$ or its constant renormalised
value can be obtained from a measurement of the frequency of the purely
orbital motion of the trapped $\phi-\chi$ pair:
\begin{equation}\label{ins.100}
  \frac{1}{2}M_{eff}(R)=\frac{B_{eff}(R)}{\Omega(R)} \;\;,
\end{equation}
where it is assumed that $B_{eff}(R)$ is already known from (\ref{ins.70})
and $\Omega(R)$ has to be measured in numerical simulation. Alternative
analytical approach is to take $\Theta$ as a geodesic coordinate
and try to find such a value of $\Omega=\dot{\Theta}$ for which
Eqs.(\ref{app.60}) posses a uniquae solution.

\subsection{ Interactions of unit vortices of a given type. }

It this subsection we would like to investigate short range interactions
of say $\phi$-vortices when there are no $\chi$-vortices or their
influence can be neglected because they are very distant.
We put $\chi=c_{2}$ and $n_{1}=n\;,\;\sigma_{1}=+1$ in this paragraph.
The configuration of n vortices splitting from their coincident position
is of the form
\begin{equation}\label{ii.10}
  \phi(r,\theta)=cF(r)[1+f(r,\theta)]e^{in\theta+i\alpha(r,\theta)}\;\;,\;\;
  \vec{V}_{(1)}\equiv\vec{V}=
                        \vec{e}_{\theta}V^{\theta}(r)+\vec{a}(r,\theta)\;\;,
\end{equation}
where the functions $F(r)$ and $V^{\theta}(r)$ satisfy  equations
\begin{equation}\label{ii.20}
  F'-\frac{n}{r}F+q_{1}FV_{\theta}=0\;\;,
\end{equation}
\begin{equation}\label{ii.30}
  \frac{-1}{r}\frac{\partial(rV_{\theta})}{\partial r}=
                                    q_{2}\gamma c^{2}_{2}(F^{2}-1) \;\;,
\end{equation}
Following similar steps as in subsection 3.1 we obtain equations
fulfilled by the perturbations:
\begin{equation}\label{ii.40}
  q_{1}a_{i}=\partial_{i}\alpha+\varepsilon_{ij}\partial_{j}f \;\;,
\end{equation}
\begin{equation}\label{ii.50}
  \nabla^{2}\alpha=0 \;\;,
\end{equation}
\begin{equation}\label{ii.60}
  \nabla^{2}f=\bar{\gamma}F^{4}f \;\;
\end{equation}
and solve them by an Ansatz
\begin{equation}\label{ii.70}
  f(r,\theta)=g(r)[\lambda_{1}cos(n\theta)+\lambda_{2}sin(n\theta)] \;\;,
\end{equation}
\begin{equation}\label{ii.80}
  \alpha(r,\theta)=\frac{-1}{r^{n}}[\lambda_{2}cos(n\theta)-
                                   \lambda_{1}sin(n\theta)] \;\;,
\end{equation}
where $g(r)$ is a solution of equation
\begin{equation}\label{ii.90}
  \triangle_{n}g(r)=\bar{\gamma}F^{4}(r)g(r) \;\;.
\end{equation}
$g(r)$ is normalised so that $g(r)\sim \frac{-1}{r}$ for small $r$.
We have taken the n-th term in the Fourier transform because it
corresponds to a uniform spliting of the n vortices from a coincident
position
\begin{equation}\label{ii.100}
     \phi \sim (z^{n}-\lambda)\;\;,\;\;\lambda=\lambda_{1}+i\lambda_{2} \;\;.
\end{equation}
The positions of the n vortices are n-th order roots of $\lambda$,
which we denote by $Re^{i\Theta+i\frac{2\pi}{n}k}$,$k=0,...,(n-1)$.
In the case of only $\phi$-vortices the effective Lagrangean reduces to
\begin{equation}\label{ii.110}
   L_{eff}=\int\;d^{2}x\;\mid\partial_{t}\phi\mid^{2}\;+\;\delta L_{eff}=
           \pi c^{2}_{1}\int\;rdr\;F^{2}g^{2}
           (\dot{\lambda}_{1}^{2}+\dot{\lambda}_{2}^{2})
                                                 \;+\;\delta L_{eff} \;\;.
\end{equation}
When we take into account that $\lambda=R^{n}e^{in\Theta}$ we can rewrite
the Lagrangean as
\begin{equation}\label{ii.120}
  L_{eff}=\frac{1}{2}M_{eff}(\dot{\lambda}\dot{\lambda}^{\star})
          +\delta L_{eff}=
          \frac{1}{2}M_{eff}R^{2(n-1)}(\dot{R}^{2}+R^{2}\dot{\Theta}^{2})+
          \delta L_{eff} \;\;.
\end{equation}
If we neglected $\delta L_{eff}$ in (\ref{ii.110}) it would follow that
in the head-on collision $\lambda$ turns
to $-\lambda$ and it is clear from (\ref{ii.100}) that it means
scattering by an angle of $\frac{\pi}{n}$. The configuration of n vortices
shrinks to a coicident position and than reappears but rotated by an angle
of $\frac{\pi}{n}$ with respect to the initial one.

   Let us analyse corrections due to $\delta L_{eff}$. First we make
a hypothesis that $\lambda$ is a geodesic
coordinate or $\ddot{\lambda}=0$ for values of $\lambda$.
The "fields" are regular as functions of $\lambda$
in $\lambda=0$. By regularity we mean finiteness and single-valuedness.
The "$\delta$-fields" are defined as linear in $\dot{\lambda}$ and we take
them as a series in the powers of $\lambda$ (components
$\lambda_{1},\lambda_{2}$). If we substitute such "fields" and
"$\delta$-fields" to the field equations (\ref{app.50}) we will find that
because "fields" are regular the "$\delta$-fields" can also be taken
self-consistently as regular in $\lambda=0$. In the limit
$\lambda\rightarrow 0$ we will obtain equations which are linear in
$\dot{\lambda}_{1}$,$\dot{\lambda}_{2}$ and in the values of
"$\delta$-fields" at $\lambda=0$. If there is a solution to these equations
we can use it to calculate $\delta L_{eff}$ and because
"$\delta$-fields" are regular at $\lambda=0$ correction to the effective
Lagrangean amounts only to renormalisation of $M_{eff}$. If there is no
solution it will mean that the initial assumption $\ddot{\lambda}=0$
was wrong and we have to look for other candidates for geodesic coordinates.

   Following argument can restrict the set of acceptable candidates. Let us
take the head-on collision of two $\phi$-vortices. Initially they are coming
to the center of mass along x-axis. Such a configuration is invariant
under succesive charge conjugation and reflection with respect to the
y-axis. Since our theory is CP-invariant the time evolution has to preserve
this symmetry of the initial configuration, so if the zeros of the Higgs
field pass through the center of mass there is possible only forward
scattering or right-angle scattering. For the right-angle scattering
$\lambda_{i}$-s are good geodesic coordinates but the forward scattering
is well described by the cartesian coordinates of vortices (zeros of the
Higgs field). If $\ddot\lambda=0$ leads to contradiction that means that we
will have to try with $R,\Theta$.

     Since these calculations are a fairly nontrivial problem we will only
conclude that only forward or $\frac{\pi}{n}$
scattering are possible in the head-on collision of vortices of the same type.
Knowledge of the exact form of $L_{eff}$ could be useful to effective
quantisation of the model and to investigations of its thermodynamics
\cite{shah}.

\section{Dual formulation}

   In this section we derive dual formulation for mutually interacting
vortices, following similar steps as in \cite{kimlee} for ordinary
Chern-Simons-Higgs system. There are two reasons for it. We would like
to show that the dual transformation can be in a natural way generalised
to the systems with more complicated Chern-Simons terms. Second and more
important - it explains why the sign of statistical interaction is inverse
to what could be expected from calculation of naive Aharonov-Bohm phase.
We will use Lagrangean (\ref{1.10}) with extra couplings to external
currents and external field
\begin{eqnarray}\label{d.10}
 L&=&\kappa\varepsilon^{\mu\nu\lambda}v^{(1)}_{\mu}\partial_{\nu}
             v^{(2)}_{\lambda}-V(f_{1},f_{2})+                  \nonumber\\
  & &\sum_{I=1}^{2}[\; \frac{1}{2}(\partial_{\mu}f_{I})^{2}+
         \frac{1}{2}f_{I}^{2}(\partial_{\mu}\omega_{I}-q_{I}v_{\mu}^{(I)}-
         e_{I}A_{\mu}^{ext})^{2}+v_{\mu}^{(I)}J^{\mu}_{(I)} \;] \;\;,
\end{eqnarray}
where $A_{\mu}^{ext}$ is an external field and $J^{\mu}_{(I)}$ are external
currents. We have rewritten Higgs fields as
   \begin{equation}\label{d.20}
     \phi=\frac{1}{\sqrt{2}}f_{1}e^{i\omega_{1}} \;\;\;,\;\;\;
     \chi=\frac{1}{\sqrt{2}}f_{2}e^{i\omega_{2}} \;\;\;.
   \end{equation}
The partition function is
   \begin{equation}\label{d.30}
     Z=\int\prod_{I}[df_{I}][d\omega_{I}][dv^{\mu}_{I}]\prod_{x}f_{I}(x)
                                           \;exp\;i\int d^{3}x\;L   \;\;\;.
   \end{equation}
Phases of the Higgs fields can be split into multivalued and regular parts
   \begin{equation}\label{d.40}
     \omega_{I}(x)=\bar{\omega}_{I}(x)+\eta_{I}(x) \;\;,
   \end{equation}
where $\bar{\omega}$-s are given by
   \begin{equation}\label{d.50}
     \bar{\omega}_{I}(t,\vec{r})=\sum_{p_{I}}\sigma_{p_{I}}
                                      Arg(\vec{r}-\vec{R}_{p_{I}}(t))\;\;.
   \end{equation}
$\vec{R}_{p_{I}}$ are positions of unit vortices $(\sigma_{p_{I}}=1)$ and
unit antivortices $(\sigma_{p_{I}}=-1)$ of the type "I". It is covenient to
construct out of $\bar{\omega}$ vortex currents
   \begin{equation}\label{d.60}
     K^{\mu}_{I}(x)=\frac{1}{2\pi}\varepsilon^{\mu\nu\lambda}
            \partial_{\nu}\partial_{\lambda}\bar{\omega}_{I}=
            \sum_{p_{I}}\sigma_{p_{I}}
            \int d\tau\frac{dR_{p_{I}}^{\mu}(\tau)}{d\tau}
                                           \delta[x-R_{p_{I}}(\tau)]
   \end{equation}
in a covariant fasion. By definition $K$ satisfies the conservation law:
$\partial_{\mu}K^{\mu}=0\;$. Integration over $\bar{\omega}$ can be replaced
by integration over vortex worldlines
   \begin{equation}\label{d.70}
     [d\omega]=[d\bar{\omega}][d\eta]=[dR_{p_{I}}^{\mu}][d\eta] \;\;.
   \end{equation}
The Jacobian $\prod_{I}\prod_{x}f_{I}(x)$ can be removed by introducing pair
of auxillary fields $C_{I}^{\mu}$:
\begin{eqnarray}\label{d.80}
\prod_{I}\prod_{x}&f_{I}(x)\;exp\;i\int d^{3}x\sum_{I}f_{I}^{2}
       (\partial_{\mu}\omega_{I}-q_{I}v_{\mu}^{I}-e_{I}A_{\mu}^{ext})^{2}=
                                                                   \nonumber\\
      &\int\prod_{I}[dC_{I}^{\mu}]\;exp\;i\int d^{3}x\sum_{I}
       [ -\frac{1}{2f_{I}^{2}}C^{I}_{\mu}C_{I}^{\mu}+
       C^{\mu}_{I}(\partial_{\mu}\bar{\omega}_{I}+\partial_{\mu}\eta_{I}-
                               q_{I}v^{I}_{\mu}-e_{I}A^{ext}_{\mu})]  &\;\;.
\end{eqnarray}
Integration over $\eta$-s will introduce
$\prod_{I}\delta(\partial_{\mu}C^{\mu}_{I})$ to the path integral measure.
These $\delta$-functions can be removed by introducing a pair of dual
gauge fields $H^{\mu}_{I}$:
   \begin{equation}\label{d.90}
     \int\prod_{I}[dC^{\mu}_{I}]\delta(\partial_{\mu}C^{\mu}_{I})=
     \int\prod_{I}[dC^{\mu}_{I}][dH^{\mu}_{I}]
               \delta(C^{\mu}_{I}-\frac{1}{2\pi q_{I}}
     \varepsilon^{\mu\nu\lambda}\partial_{\nu}H^{I}_{\lambda}) \;\;
   \end{equation}
and integrating over auxillary fields $C^{(I)}_{\mu}$.
Now the vortex currents can be introduced by the identity
   \begin{equation}\label{d.100}
     \int d^{3}x\;\frac{1}{2\pi q_{I}}\varepsilon^{\mu\nu\lambda}
       \partial_{\mu}\bar{\omega}_{I}\partial_{\nu}H^{I}_{\lambda}=
     \frac{1}{q_{I}}\int d^{3}x\;K^{\mu}_{I}H^{I}_{\mu} \;\;,
   \end{equation}
where integration by parts has been done and we have made use of the
definition of $K^{\mu}_{I}$ , see (\ref{d.60}). The present intermediate
form of the Lagrangean reads
\begin{eqnarray}\label{d.110}
L&=&\kappa\varepsilon^{\mu\nu\lambda}v_{\mu}^{(1)}\partial_{\nu}
          v_{\lambda}^{(2)}+\sum_{I}\frac{1}{2\pi}v_{\mu}^{I}
          \varepsilon^{\mu\nu\lambda}\partial_{\nu}H_{\lambda}^{I} \nonumber\\
 & &    +\sum_{I}[-\frac{1}{16\pi^{2}f_{I}^{2}q_{I}^{2}}
        H^{I}_{\mu\nu}H^{\mu\nu}_{I}+\frac{1}{q_{I}}H^{I}_{\mu}K^{\mu}_{I}+
        \frac{1}{2}(\partial_{\mu}f_{I})^{2}-
        \frac{e_{I}}{4\pi q_{I}}\varepsilon^{\mu\nu\lambda}H^{I}_{\mu}
        F^{ext}_{\nu\lambda}+v^{I}_{\mu}J^{\mu}_{I}]               \nonumber\\
 & &     -V(f_{1},f_{2})
\end{eqnarray}
and the integration measure
    \begin{equation}\label{d.120}
      \prod_{I}[df_{I}][dR^{\mu}_{p_{I}}][dH^{\mu}_{I}][dv^{\mu}_{I}] \;\;.
    \end{equation}
  We would like to remove the gauge fields $v^{\mu}_{I}$ from the Lagrangean.
Let us take a look at their classical field equations following from
(\ref{d.110}):
    \begin{equation}\label{d.130}
      \kappa\varepsilon^{\mu\nu\lambda}\partial_{\nu}v_{\lambda}^{(2)}+
      \frac{1}{2\pi}\varepsilon^{\mu\nu\lambda}\partial_{\nu}H^{(1)}_{\lambda}
      +J^{\mu}_{(1)}=0 \;\;,
    \end{equation}
    \begin{equation}\label{d.140}
      \kappa\varepsilon^{\mu\nu\lambda}\partial_{\nu}v_{\lambda}^{(1)}+
      \frac{1}{2\pi}\varepsilon^{\mu\nu\lambda}\partial_{\nu}H^{(2)}_{\lambda}
      +J^{\mu}_{(2)}=0 \;\;,
    \end{equation}
Motivated by these equations we write
    \begin{equation}\label{d.150}
      v_{\mu}^{(1)}=-\frac{1}{2\pi\kappa}H_{\mu}^{(2)}+G_{\mu}^{(1)} \;\;\;,
    \;\;\; v_{\mu}^{(2)}=-\frac{1}{2\pi\kappa}H_{\mu}^{(1)}+G_{\mu}^{(2)} \;\;,
    \end{equation}
where $G_{\mu}^{(I)}$ are extra fields due to the presence of external
currents and containing quantum fluctuations around classical solution.
Thus finally the partition function reads
    \begin{equation}\label{d.170}
      Z=\int\prod_{I}[df_{I}][dR^{\mu}_{p_{I}}][dH^{\mu}_{I}][dG^{\mu}_{I}]
        \;exp\;i\int d^{3}x\;L_{D} \;\;,
    \end{equation}
where the dual Lagrangean is
\begin{eqnarray}\label{d.180}
L_{D}&=& -\frac{1}{4\pi^{2}\kappa}\varepsilon^{\mu\nu\lambda}
              H^{(1)}_{\mu}\partial_{\nu}H^{(2)}_{\lambda}
              +\sum_{I}[-\frac{1}{16\pi^{2}f_{I}^{2}q_{I}^{2}}
                                                H^{I}_{\mu\nu}H^{\mu\nu}_{I}
              +\frac{1}{q_{I}}H_{\mu}^{I}K^{\mu}_{I}]               \nonumber\\
     & &     +\sum_{I}
             [\frac{1}{2}(\partial_{\mu}f_{I})^{2}]-V(f_{1},f_{2})] \nonumber\\
     & &     +\kappa\varepsilon^{\mu\nu\lambda}
             G^{(1)}_{\mu}\partial_{\nu}G^{(2)}_{\lambda}          \nonumber\\
     & &     \sum_{I}[-\frac{e_{I}}{4\pi q_{I}}\varepsilon^{\mu\nu\lambda}
             H_{\mu}^{I}F^{ext}_{\nu\lambda}+G^{I}_{\mu}J^{\mu}_{I}]
             -\frac{1}{2\pi\kappa}H_{\mu}^{(2)}J^{\mu}_{(1)}
             -\frac{1}{2\pi\kappa}H_{\mu}^{(1)}J^{\mu}_{(2)}  \;\;.
\end{eqnarray}
 When there is no external current the fields $G_{\mu}^{I}$ decouple
and can be integrated out giving contribution to the normalisation factor.
If in addition $F^{ext}_{\mu\nu}=0$ we are left only with the first two lines
of the above dual Lagrangean.

 Vortex current couple to the dual fields $H^{\mu}$. Since the
Chern-Simons term for the dual fields has an opposite sign to that
for the original gauge fields $v^{\mu}_{I}$ it explains why the statistical
interaction term in the effective Lagrangean (\ref{1.135}) has an opposite
sign to that
expected from the values of vortex fluxes and charges. The dual Aharonov-Bohm
interaction between vortices is mediated by the dual fields and it gives rise
to the correct value of the statistical interaction.

\section{Conclusions}

   We have made an analysis of interactions of self-dual Chern-Simons
vortices in the limit of very large and very small separations. We have
shown the existence of mutual statistical interaction between vortices
of different types in $[U(1)]^{2}$ model but the results can be easily
generalised to the general $[U(1)]^{N}$ theory. The sign of the statistical
interaction is inverse to expectations based on ordinary Aharonov-Bohm effect.
Thats why we have derived dual formulation of the system in which
it is clear that vortices interact via dual gauge field with the sign
of the mutual Chern-Simons term inverse to that in the original formulation.

   We have not attempted calculating corrections to the standard adiabatic
approximation but a possible method how it could be done was discussed.
If the corrections are only quantitative in nature the qualitative picture
of short range interactions obtained in ordinary adiabatic approximation
remains unchanged. In the head on collision of vortices of the same type
we should observe right-angle scattering. For vortices of different types
at large separations dual Aharonov-Bohm effect is observed but when
their cores overlap they behave like charged particles crossing magnetic flux.

  The analysis of the short range interactions of vortices in
Chern-Simons-Higgs systems presented in both this article and in
\cite{kimlee,mis} shows the possibility of periodic solutions very much like
bound states of vortices. The semiclassical quantisation of these solutions
\cite{jackiw} can give rise to some discrete spectra of energy. The spectra
can be expected as an additional quantum effect to the statistical
interaction due to the finite width of vortices. The special effects
of the short range interactions could be expected to vanish in the limit
of vanishing thickness of vortices but we know from the studies
of the string limit for vortices in Abelian Higgs model \cite{gervais}
that even when classical vortices become very thin the quantum fluctuations
cause that they preserve nonzero effective thickness. So it is possible
that a classical vortex of finite width is a better zero order approximation
to the full quantum theory \cite{sailer}.

   The other topic worth of detailed study is the possibility of existence
of ordinary fractional statistics in the apparently only mutually interacting
system. The pair of $\phi$ and $\chi$ vortices can form a composite thanks
to the magnetic trapping. If the potential of the model $V(\phi,\chi)$ were
slightly deformed in such a way that it would prefer energetically
overlapping of the $\phi$-vortices and $\chi$-vortices but it would discourage
vortices of the same type to overlap, than we would expect vortices of
different types to form true stable bound states. If we wanted this bound
states to be composed of exactly one vortex of each type we would have
to make a repulsion of the species of the same kind to be stronger than
attraction of vortices of different types. Such a multivortex system in
a sufficiently low temperature would be a gas of such anyonic bound states.
In a higher temperature the average kinetic energy of the anyons could be
large enough to split them into particular mutually interacting vortices.
Thus we can construct a system with two phases: an anyonic one and a phase
with mutual statistics. The composed anyons might have an interesting
internal structure. If the corrections to the Higgs potential do not
change to much the interactions patterns at small separations the two vortices
will feel both the charge-flux interaction and an oscilatory type interaction
due to the atractive properties of the Higgs potential. The phase-diagram
of the system could be even more complicated if the Higgs potential itself
depended on temperature. We think all these topics to be worth of further
investigation.

    $Aknowledgement.$ I would like to thank Prof.H.Arod\'z for discussion
and reading of the manuscript.

\end{document}